\documentclass[preprint,floats,aps]{revtex4}
\usepackage{dcolumn}% Align table columns on decimal point
\usepackage{bm}% bold math
\usepackage{graphicx}
\begin{document}
%\baselinestretch{5}
%\tightenlines
\topmargin=-0.3cm
\title{A Novel Dynamical Approach To Relativistic Heavy Ion Collisions}
%\vspace{0.1in}
\author{Dai-Mei Zhou$^{1-2}$, S. Terranova$^1$ and A. Bonasera$^1$
 \footnote{Email: bonasera@lns.infn.it}}
\affiliation{
$^1$  Laboratorio Nazionale del Sud, Istituto Nazionale Di Fisica Nucleare,
      Via S. Sofia 44, I-95123 Catania, Italy \\
$^2$  Institute of Particle Physics, Huazhong Normal University, Wuhan,
      430079 China
}
%\maketitle
\begin{abstract}
A transport model for ultra-relativistic nucleus-nucleus
collisions based on the mean free path approach is proposed.  The method
is manifestly Lorentz invariant.
We discuss some calculations for pp and AA collisions and compare to a previously proposed transport
model and to data.  We demonstrate that our approach gives a different impact parameter distribution already
 in pp collisions as compared to the previous one.  The role of hadronization times is discussed.
  Comparison to data is reasonable and the model
 can be easily modified to take into account genuine many body effects and quantum statistics similarly
 to low energy heavy ion collisions.\\
\noindent{PACS numbers: 25.75.-q, 12.38.Mh, 24.10.Lx}
\end{abstract}
%\vspace{0.1in}
\maketitle

%\begin{figure}[ht]
%\centerline{\hspace{-0.5in}
%\includegraphics{file=pau_pi.EPS,width=4.5in,height=2.0in,angle=0}}
%\vspace{0.2in}
%\caption{Rapidity (left panel) and transverse momentum (right panel)
% distributions of $\pi^+$ in $Pb+Pb$ collisions at $\sqrt{s_{nn}}$=5500 GeV.}
%\label{pau_pi}
%\end{figure}

Relativistic heavy ion collisions (RHIC) offer a unique tool to explore matter states never
 explored before.  In particular at high energy densities a transition from nuclear matter to the quark
gluon plasma (QGP) is expected.  Because of the high complexity of
the problem two different worlds, traditionally known as particle
and nuclear physicists, are joining their efforts to gather
important information on the transition to the QGP.  This implies
that many techniques and physics developed in the two fields are
merged to have a complete picture of the AA collisions.  In fact,
in these collisions we can disentangle the necessary input coming
from elementary collisions with creation of new particles from the
dynamics of quarks and gluons (partons), and the time evolution of
the many body systems which is extremely complicated due to the
initial number of particles in AA collisions
 which largely increases during the time evolution depending on the beam energy.  It is of course necessary
to have both aspects under control to understand the complex physics of the process.  The purpose
of this paper is to propose a new method of solution of the semiclassical transport equation which should
 describe (in some approximations which we will not discuss here \cite{aich,gei}) the many body part of the
process.  The kinetic approach used so far is essentially a generalization of transport codes proposed
for low energy heavy ion collisions \cite{cug,bert,aich} (known as  Boltzmann Uehling Uehlenbeck (BUU), Vlasov
(VUU)/ Landau (LV)).  It is based
 on  following the time evolution of each particle with a collision
occurring  if two of them come to a
closest distance less than or equal to $b_{max}=$ $\displaystyle{\sqrt
{\sigma_{tot}/\pi}}$. Here $\sigma_{tot}$ refers to the total cross section.   The method we discuss in
this work is known as Boltzmann Nordheim Vlasov (BNV) approach at low energies \cite{aldo}.  It
is based on the concept of the mean free path approach and we will discuss it in detail in the following
section.  We would like to stress here that even though the two approaches might be
 comparable as far as
average quantities are concerned, their detailed description of
the dynamics is quite different.  In fact in the first  method
(I), a particle
 is treated as a black sphere and if a second particle comes within a radius
defined from the elementary cross section, then they collide.
 In the method (II) discussed in this work, a
probability of collision is defined (see below) and particles can collide according to this probability.  Thus,
for instance in pp collisions if N particles are produced at a given energy, within method I, N particles will
be produced for $0<b<b_{max}$, while in method II N(b) particles are produced at each b and N particles are
obtained after averaging over b.  Thus even though the two methods might
give the same average number their
 b dependence is quite different and this might be important especially when one searches for very violent
events, i.e. events at very high multiplicities presumably obtained in very central collisions.

    For the discussion in this work we will base our considerations on a parton cascade model
for ultra-relativistic heavy ion collision \cite{gei}
generally composed of the parton initial state, the parton evolution, the
hadronization, and the hadron evolution\cite{sa1}.
  There are two ways of creating the
parton initial state. The parton initial state was composed of
partons from the mini-jets production in nucleus-nucleus collision
and the HIJING multiple mini-jet generator \cite{wang} was
specified in \cite{bin,gyu}. The parton initial state in
\cite{boer,bass} was created via probability distributions first
for the spatial and momentum coordinates of nucleons in colliding
nuclei and then for the flavor and spatial and momentum
coordinates of partons in nucleons. Our parton cascade model
follows the former way, however, the JPCIAE multiple mini-jet
generator \cite{sa2} is used instead of HIJING.  The JPCIAE
multiple mini-jet generator for ultra-relativistic nucleus-nucleus
collision is based on PYTHIA \cite{sjo1} which is a well known
event generator for hadron-hadron collisions.  Thus in this work we
will study the two different approaches starting from the same
numerical code which has been in case II opportunely modified as
discussed below.   An important ingredient of the approach is the
hadronization time (ht).  This is the time it takes for partons
created after a collision to coalesce and  form a new hadron.  We
will perform some calculations assuming a ht equal to zero and
compare to calculations where the ht is nonzero. In the latter
case any finite ht value will suppress secondary hadron-hadron
collisions because, at high beam energies, the ht is time
elongated, thus it might be larger than the reaction time. This is
a very important effect because it suggests that we can have a
finite hot region of quarks and gluons which will not coalesce
into hadrons for a ht time boosted to the nucleus-nucleus CM
frame.

 Quantum statistics (i.e. Pauli and Bose statistics) are not included at present
but they have been considered explicitly in  \cite{bert1}
 for Bose and \cite{sa04} for Fermi statistics.
In future work we plan to include the different statistics and
the influence of many particle
 collisions \cite{aldo} which could be very important because of the reached high densities.

 \section{Formalism}

We follow the mean free path method as discussed
in\cite{aldo} and modified to include relativistic effects.  Briefly, for each event, at each time
 step dt and for each ion $i$ we search
for the closest ion $j(i)$ in phase space i.e. we define the quantity
 \begin{equation}
\Xi_{ij}=\frac{r_{ij}}{v_{ij}dt}
\end{equation}

where ${r_{ij}}, {v_{ij}}$ and dt are the relative distances, velocities (relativistic) and $dt$ is
the time step used in the calculations.
Define a collision probability as:
 \begin{equation}
\Pi=\frac{v_{ij}dt}{\lambda}=\sigma\rho(r_i)v_{ij}dt ,
%\label{lv3}
\end{equation}
where $\rho(r_i)$ is the local density calculated at each time step.
  Note that the quantities
defined above are scalars and it is quite easy to show that they are
 Lorentz invariant.  For instance
in eq. (1) we have a distance  (${r_{ij}}$)  divided by a distance (${v_{ij}dt}$).

The physical meaning of the equation (1) is simple.
  We search for particles that are close
in coordinate space  and far away in velocity space, i.e.
particles with opposite momenta.  For instance at RHIC energies we
have a relative velocity of the order of the speed of light c for
particles of the target (T) and projectile (P)  respectively and
zero for both particles belonging to T or (P).  The latter
particles are automatically excluded from the first condition.
Once the two closest particles have been found, we calculate the
local density knowing the relative distance and the number of
particles near the colliding couple. From the relative energy and
the particles type we know the elementary cross sections and thus
the probability eq.(2).  A random number is taken and if smaller
than the calculated probability, a collision occurs and it is
modeled by Pythia if $\sqrt(s) >4 GeV$ or some other value
 which we will specify when fitting pp data.
   Because of the probabilistic
nature of the process discussed above we can have events where particles are very close and they do
not collide, while it can happen that for smaller densities a collision occurs with the small probability
 $\Pi$.  The time step is chosen such that the probability is always small compared to 1 and to minimize
the CPU time.  Once these conditions are fulfilled a further decrease of the  time step dt will not change
 the results but will increase linearly the CPU time.  Notice that in our approach we have to pay the
price of largely CPU time consuming calculations, proportional to $N(t)^2$ for zero ht, where N(t) is the number of
particles at time t.  Such a proportionality with $N(t)$ decreases for finite ht.
 This calculation time might become prohibitive at LHC energies where many particles
are produced.  In the following we will show if it is worth or not
to follow this approach or stick to the previous one.  A
discussion of the implementation of the alternative approach is
not given here since it has been largely discussed in the
literature \cite{gei,sa1}. In both approaches the partonic system
is  hadronized by JETSET \cite{sjo1} after some ht.   As we
discussed above the role of ht is crucial.  In fact, if such a
time is small the partons hadronize and can collide again.  In the
next section we will study the effects of rescatterings when the
ht is zero and/or very small.
   But if the ht is finite, then, also because of Lorentz dilatations hadrons cannot be formed
and there cannot be other hadron-hadron (hh)  collisions for the ht duration.  In the limit of large ht all the secondary collisions are suppressed and our approach reduces to Pythia folded with the number of first
chance nucleon-nucleon collisions.
  We will also study the effects of finite ht effects.  We
would like to stress at this stage, that even though hadrons are not formed the partons could still collide.
For illustration purposes we will neglect those collisions in our calculations, however we would like to
notice the following important points.

i) At the partons stage , collisions might occur among partons.  One simple
way to implement this would be to scale $\it{down}$ the hh cross sections by the number of valence quarks (or of participant
 partons).  But in this case the local density should be scaled $\it{up}$ of the same number which in turn gives the same
mean free path of hadrons, cf.. eq.(2).  Thus,  in this approximation, we expect that including parton-parton collisions will
have the same effects as having a ht equal to zero, because the hh and parton-parton mean free paths are the same as discussed above.

ii) If at some stage we would decide to include parton-parton collisions then as a first step we have to check
that our approach coupled to Pythia fits the experimental pp collisions for instance.  This might lead to
a redefinition of the Pythia parameters.  In other words, even though we are not formally including collisions
at the parton level, those might be implicitly included in the parametrization of the elementary collisions.

 Important differences for the collisions (thus for the dynamics) are due to the statistics.
 In fact, if we have ht equal to zero, our
system is made up essentially of pions, while for finite ht we have quarks and gluons.
In the hadronic state essentially Bose
 Einstein statistics applies while in the partonic stage, both Bose and Fermi statistics apply.
 This, we feel will make an important
distinction and will be discussed in detail in following works.

The mean free path method discussed above has been studied in detail at low energies and it has been
shown to solve the Boltzmann eq. in the cases
 where an analytical solution is known \cite{aldo}.  Here we have
generalized the approach to keep into account relativistic effects.  The particles move on straight lines
during collisions since we have not implemented any force yet.  For short we name the method proposed
here as Relativistic Boltzman equation (ReB).

 \section{hadronic picture}

The essential inputs of all transport approaches are the elementary cross sections.  Of course
the used models Pythia and Jetset are tailored to reproduce those elementary processes.  
 In this
paragraph we will show what is the influence of the way the collision partners are chosen and we will
fix the essential parameters which enter the calculations.  The physical picture of the collision is the following.  The colliding partners are chosen as discussed in the previous section. 
Depending on their relative energy it is decided if the collision is elastic or inelastic.  If it is inelastic, the number of produced partons and their energies and momenta are decided using Pythia.  Those partons are randomly distributed in a sphere whose diameter is given by the original positions of the colliding hadrons. The partons are evolved on straight line trajectories and they cannot collide before the ht.  After the ht the partons hadronize and the formed hadrons can collide again.  Clearly in  elementary collisions, if the ht is long enough there will not be secondary collisions, thus our approach for elementary collisions reduces essentially to Pythia.  But, on the other hand if the ht is zero or very short, the produced hadrons will be still very close in phase space and they might collide again.  In this case we  adjust the parameters of Pythia such that the elementary cross sections are still reproduced.  Notice that this is not a feature of our approach but it is common to other methods of solution of the transport equation such as JPCIAE or URQMD.  Now,  we discuss the role of the ht,
 which influences the rescatterings.
\begin{figure}[ht]
\centerline{\hspace{-0.5in}
\includegraphics[width=4.5in,clip]{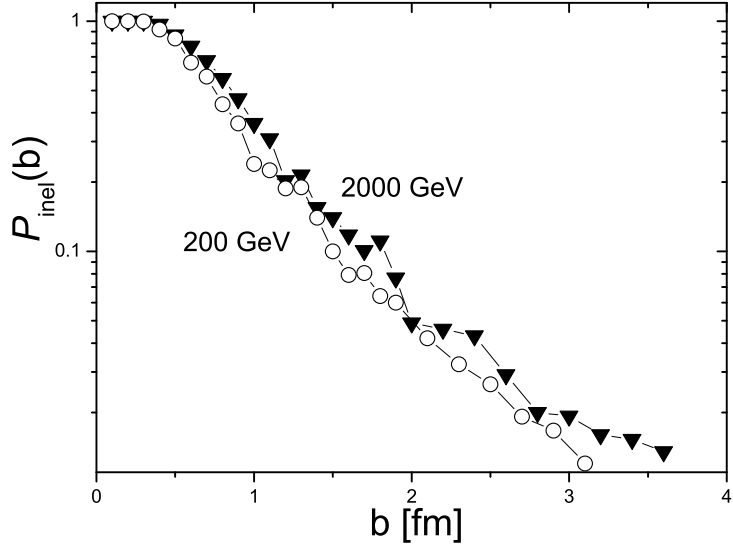}}
\vspace{0.2in} \caption{Particles production probability versus
impact parameter in pp collisions at two C.M. energies,
 200 and 2000 GeV, respectively.}
\label{probx}
\end{figure}

First, in figure (1) we show the multiplicity of produced charged particles in pp collisions
 vs. impact parameter
b.  As we discussed above since a collision is chosen in a probabilistic way there are events where
a collision occurs and events where they do not occur (apart b=0 fm where a collision strictly occurs for each
event since the density calculated from the relative distance of the 2 particles diverges).  This
 explain the b dependence decrease of the multiplicity.
 Not only, we see that the Blacker, Edger and Larger\cite{gei}
 (BEL) effect is reproduced:  for higher energies the elementary cross sections increases thus the number of
involved impact parameters increases as well.  This feature is of course contained in the original JPCIAE model
as well since the same cross section (obtained from data) is used.  However the b dependence of the multiplicity
distribution is dramatically different. In fact JPCIAE
gives a constant yield as function of b.  For sake of simplicity in this work we tune the parameters entering
 the calculations to  pp collisions at zero impact parameter. For instance in fig.2 we display some results
 for zero hadronization time.  We stress the fact that zero ht implies that at each collision the produced
particles are hadronized instantaneously, thus we could say that this approximation correspond to a
purely hadronic transport model.
\begin{figure}[ht]
\centerline{\hspace{-0.5in}
\includegraphics[width=4.5in,height=3.0in,angle=0]{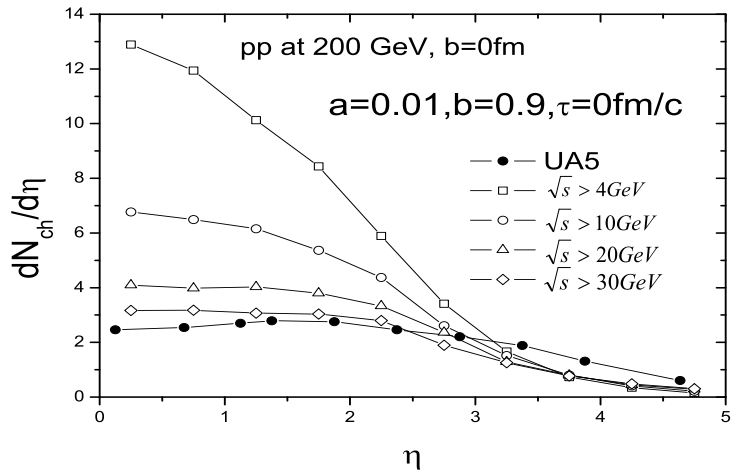}}
\vspace{0.2in} \caption{Particles production  versus
pseudo-rapidity in pp collisions at
 200 GeV C.M. energy for zero hadronization times.}
\label{fig2}
\end{figure}

In the figure (2) we show the particles multiplicities in pp collisions from REB calculations, compared
 to data (full circles) \cite{ua5} for various energy cutoffs  and parameter sets entering Pythia.
 The fact that ReB results for strictly zero ht and drastically reducing the a and b parameters (which tune
the strength of the interaction in Pythia) does not reproduce the
data is due to secondary collisions which increase the yield
especially of low pseudo-rapidity particles.  In order to improve
the agreement to data we  increased the low energy cutoff
for Pythia up to 30 GeV.  Notice however that
the energy  cutoffs have the same effect of a larger ht, in fact what
happens is that inelastic secondary collisions occurring at later
times are suppressed by our choice of 30 GeV cutoff and this has a
similar effect to increase the ht which avoids secondary
collisions at later times. However as we will show in the
following a finite ht or a large cutoff in
 the energy does not have the same effect in  AA collisions.  
 Thus we can already conclude that zero ht (or
very small ht) in our model are already excluded from pp collisions.  This become strikingly evident
 when comparing to data for Au+Au collisions at $\sqrt{s}=200 GeV$, see figure(3).   In fact the data \cite{bear} for pseudorapities
less than roughly 3 units are largely overestimated especially for kaons.  This suggests that those particles are produced in the model at
later stages of the reaction from lower energies colliding hadrons.

\begin{figure}[ht]
\centerline{\hspace{-0.5in}
\includegraphics[width=4.5in,height=3.0in,angle=0]{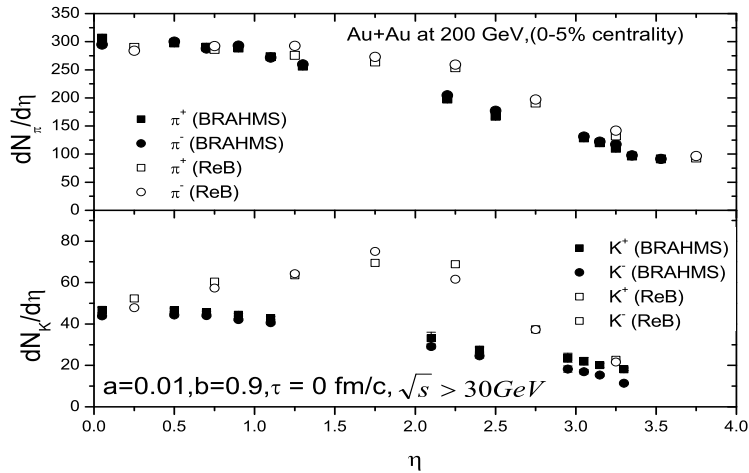}}
\vspace{0.2in} \caption{Particles production probability versus
pseudo-rapidity in Au+Au collisions at  C.M. energy
 200  GeV and for central collisions. Data for charged pions (top panel) and charged kaons (bottom panel),
 full symbols, are also included\cite{bear}.}
\label{aatau0}
\end{figure}

\section{Influence of the hadronization times}
\begin{figure}[ht]
\centerline{\hspace{-0.5in}
\includegraphics[width=4.5in,height=3.0in,angle=0]{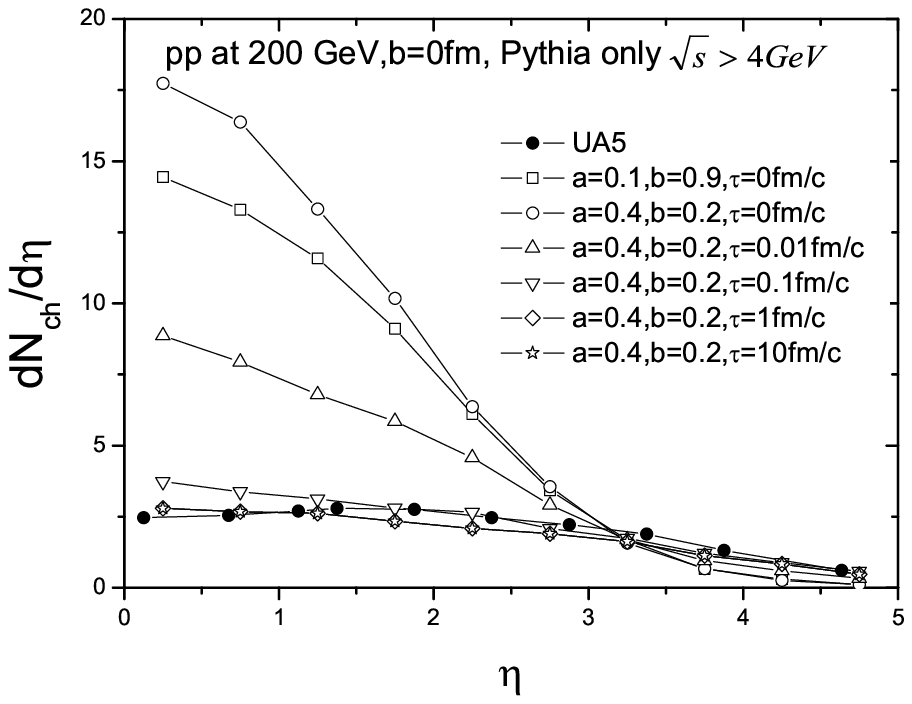}}
\vspace{0.2in} \caption{Particles production  versus
pseudo-rapidity in pp collisions at
 200 GeV C.M. energy for different hadronization times.}
\label{fig4}
\end{figure}

As we stressed above after an elementary hh collision, quarks and gluons are created which
eventually  combine to give new hadrons.  This process takes some time (ht).
In this section we will show the effects of finite ht assuming that during such a time the partons
cannot collide.  This is of course an approximation since collisions might occur among partons.
 However, especially if we have a QGP, the possibility of colliding is quite different from the
case where there is no phase transition.  Thus to simply scale the elementary hh collisions by,
for instance, the number of valence quarks could be too naive if the partons are embedded on
a plasma.  As we stressed above, the simple scaling of the cross sections goes together with a
scaling of the local density which will result in a similar mean free path of hadrons and partons,
 thus in our framework, the results including parton collisions (neglecting quantum statistics and
modified partons cross sections in a QGP) will be very similar to
the results discussed in the previous sections with ht equal to
zero.  Instead, blocking the possibility of partons inelastic
collisions, might give some hint, through a comparison to data of
the relevance of the missing dynamics during the partons stage.
But again we have to stress that maybe some parton-parton
 collisions are implicitly included in the model through the fitting of the  pp data.
\begin{figure}[ht]
\centerline{\hspace{-0.5in}
\includegraphics[width=4.5in,height=3.0in,angle=0]{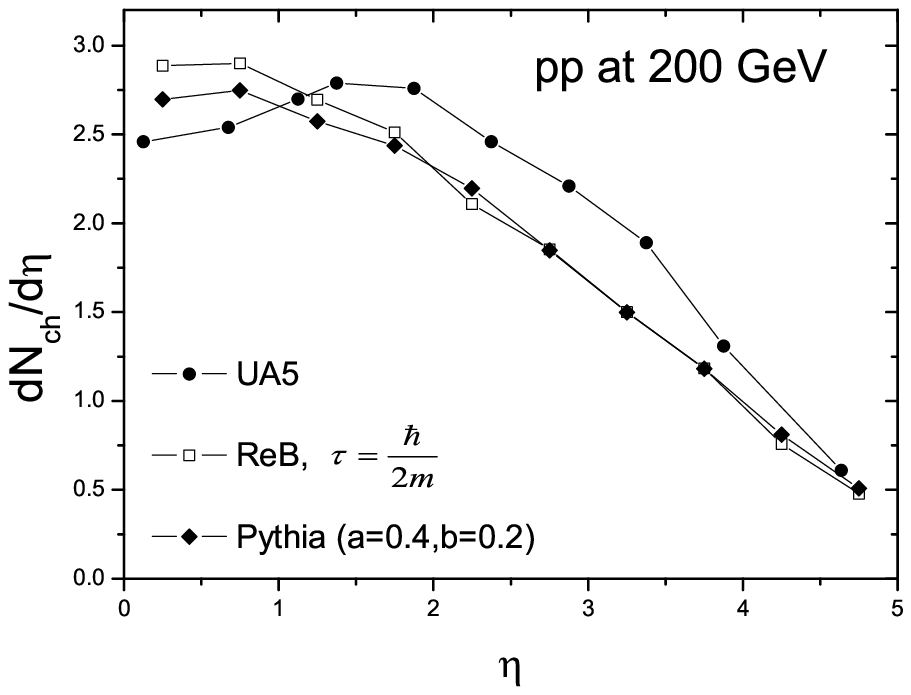}}
\vspace{0.2in} \caption{Particles production  versus
pseudo-rapidity in pp collisions at
 200 GeV C.M. energy for 'Heisenberg' hadronization times.}
\label{fig5}
\end{figure}

There is not a definite prescription on what should be the ht
times both in pp and AA collisions.  Thus we will treat it as a
free parameter, but for each choice we make we will first fit the
experimental pp data
 by tuning the Pythia parameters exactly as we did in the previous section.  In figure (4) we plot
  the pseudo-rapidity distributions compared to UA5 data \cite{ua5} (full circles).

\begin{figure}[ht]
\centerline{\hspace{-0.5in}
\includegraphics[width=5.5in,height=3.0in,angle=0]{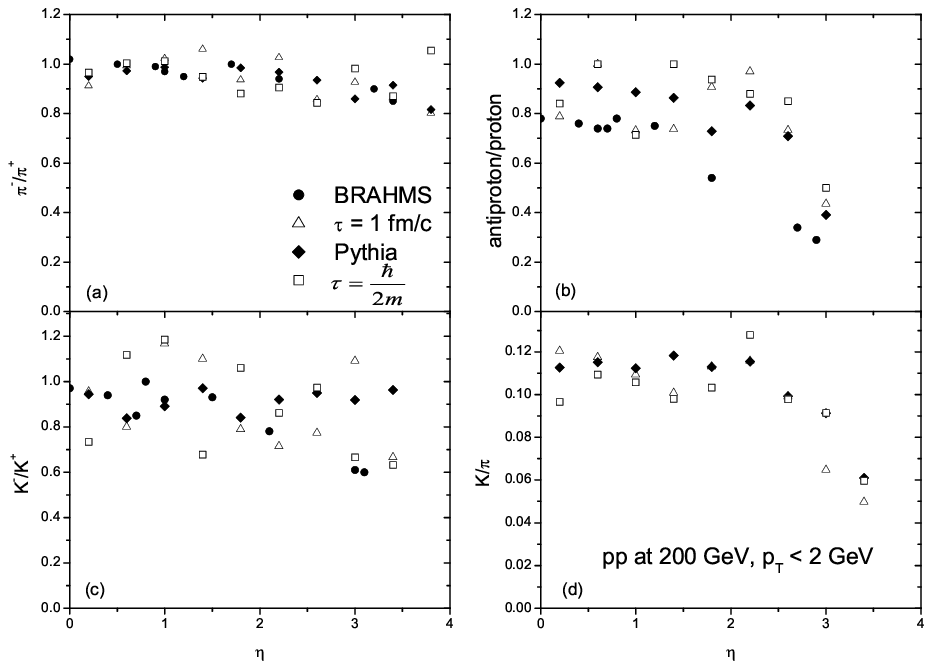}}
\vspace{0.2in} \caption{Different hadron ratios versus
pseudo-rapidity in pp collisions at
 200 GeV C.M. energy for different hadronization times.}
\label{fig6}
\end{figure}

Different ht values are displayed in the figure together with the
set of Pythia adjusted parameters.  Again a zero ht time does not
fit the pp data, while finite values of the ht do a reasonable
job.  Notice also that the results of the fits are not so
dependent on the ht if this is larger than 0.1 fm/c.   In fact
because of Lorentz time dilatation  hadronization does not occur
before the system has expanded thus suppressing secondary
collisions.  This is also demonstrated by the good agreement of
those calculations to the Pythia results with the same parameter
sets.

\begin{figure}[ht]
\centerline{\hspace{-0.5in}
\includegraphics[width=6.5in,height=4.0in,angle=0]{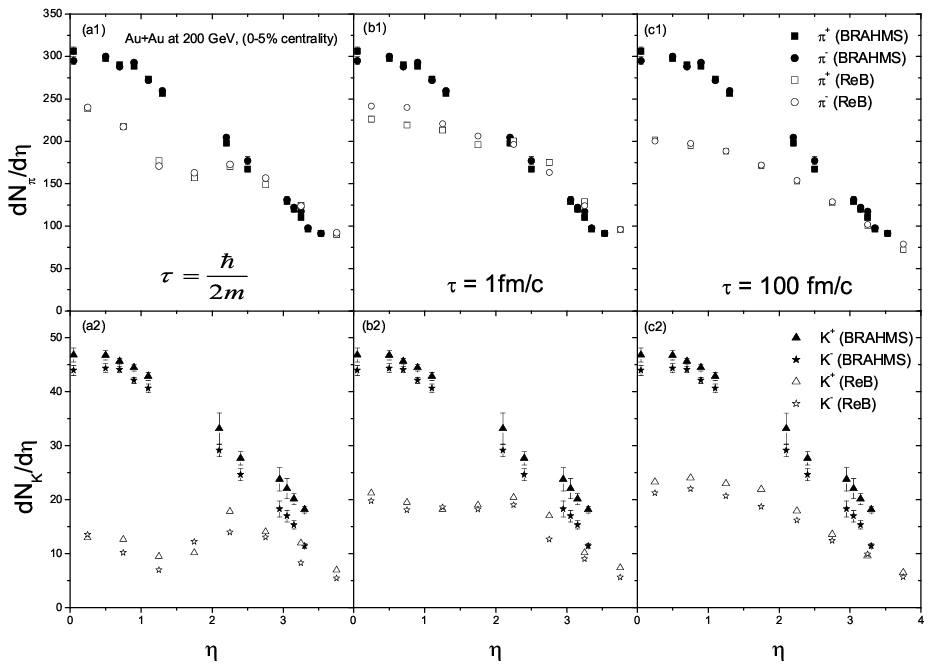}}
\vspace{0.2in} \caption{Particles production  versus
pseudo-rapidity in Au+Au collisions at
 200 GeV C.M. energy for different hadronization times.}
\label{fig7}
\end{figure}

For illustration in fig.(5) we plot the result for the same pp collisions but with a ht $\tau=\hbar/2m$, i.e.
 the Heisenberg principle, where m is the mass of the forming hadron.  This choice was done to see if there
are some differences when the ht depends on the produced particle type.

To see if different choices of the ht give noticeable differences
already in pp collisions, we plot in figure (6) various hadronic
ratios vs. rapidity at the same energy as above.  The data, when
available, is given by the full circles with error
bars\cite{brampp}. We see that our results do a similar job of the
original Pythia approach (diamonds)
 for $\pi^-/\pi^+$  and $K^-/K^+$ ratios while we do not reproduce well the $\bar{p}/p$ ratio.
 \begin{figure}[ht]
\centerline{\hspace{-0.5in}
\includegraphics[width=4.5in,height=3.0in,angle=0]{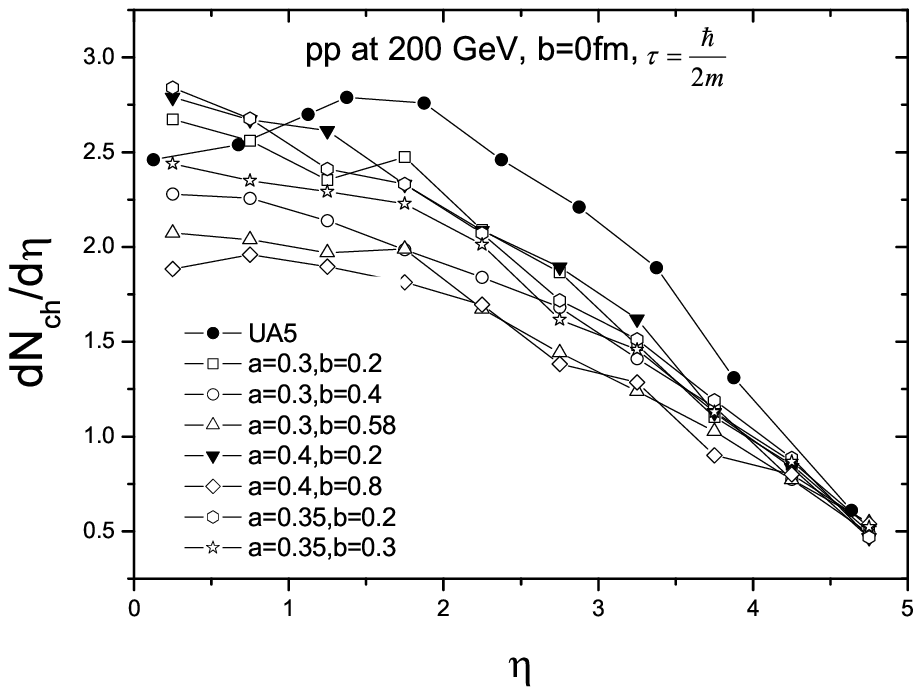}}
\vspace{0.2in} \caption{Particles production  versus
pseudo-rapidity in pp collisions at
 200 GeV C.M. energy with no cutoff in the elementary cross sections parametrization.}
\label{fig8}
\end{figure}

\begin{figure}[ht]
\centerline{\hspace{-0.5in}
\includegraphics[width=4.5in,height=3.0in,angle=0]{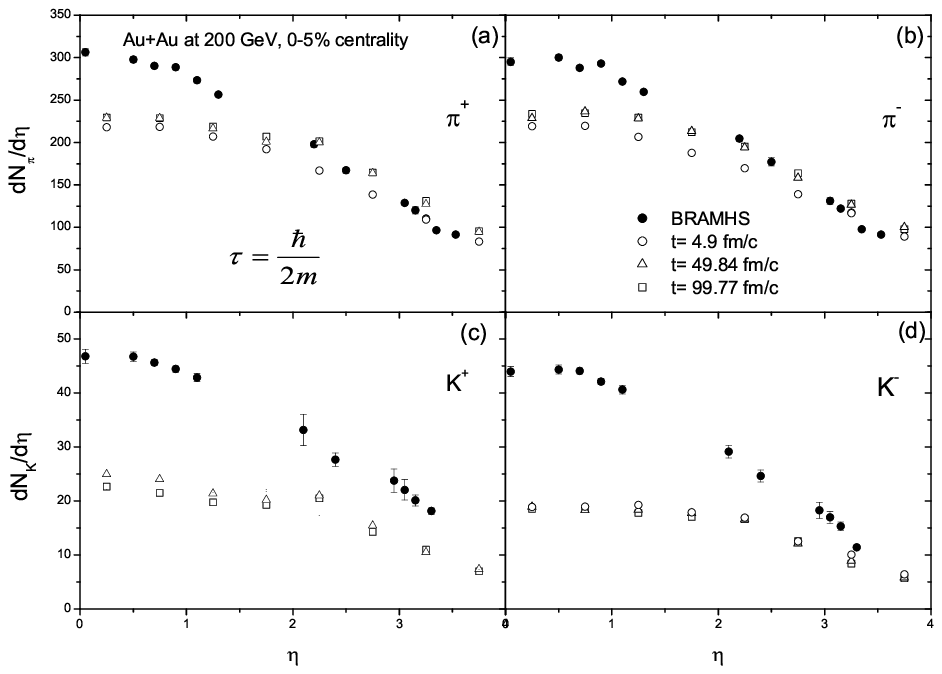}}
\vspace{0.2in} \caption{Particles production  versus
pseudo-rapidity in Au+Au collisions at
 200 GeV C.M. energy at  different  times.}
\label{fig9}
\end{figure}

Let us now turn to the Au+Au collisions at RHIC. In  the
different panels of fig.7, we display the results of the calculations (open
symbols) using different ht for pions (top panels) and kaons
(bottom panels).  The experimental data \cite{bear} is also
displayed. The discrepancy with the data is striking especially at
low pseudo-rapidity.  No noticeable differences are
 found for the two first different prescriptions for the ht.  When using a very large value of the ht (which
implies first chances collisions only)  the agreement is poor as
well.   If we compare to the result for zero ht, cf. fig.2, we
notice that we have largely improved the description of the data
especially for pions and $\eta > 2$.  However, the results for lower pseudo-rapidities are not
as good as those of fig.2. The difference among the
calculations is not only the ht but the low energy cutoff for
Pythia.  Recall that for zero ht we increased the value of the
energy cutoff to 30 GeV while for finite ht the smallest possible value for Pythia5.7 
is used, i.e. 4 GeV.  Therefore, as in the zero ht case the
low energy cutoff was responsible for decreasing the low $\eta$
yield, it could be  that the 4 GeV cutoff in Pythia is
responsible as well for the bad reproduction of the data for lower
$\eta$s.  This called for extra work of our postdocs who had to
implement a new subroutine with a parametrization of elementary
collisions data for energies below 4 GeV and above the energy for
pion production. This was finally accomplished and the final
result for finite ht is displayed in figs.8-11. In fact,
first we had to fit the pp data again, (see  one example in  fig.8)
 and then
 we compared to the Au+Au case(fig.9-11). The agreement to data is now better both for pions and
kaons production, even though we slightly overestimate the data at
small $\eta$s.  The way we have reached this result
suggests that low $\eta$ particles are produced from hh collisions
at lower energies and later times.  Also, by comparing to fig.2, one might try to use smaller values
 of the ht  to fit the data.  However, a reproduction of the data by fitting some parameter is not the purpose of this work.  In fact,  before comparing in detail to data we would like to add first some missing ingredients (quantum statistics, cooperative effect etc.).  More on this point plus a
discussion on the time development of the reaction is given in the
next section.
\begin{figure}[ht]
\centerline{\hspace{-0.5in}
\includegraphics[width=4.5in,height=3.0in,angle=0]{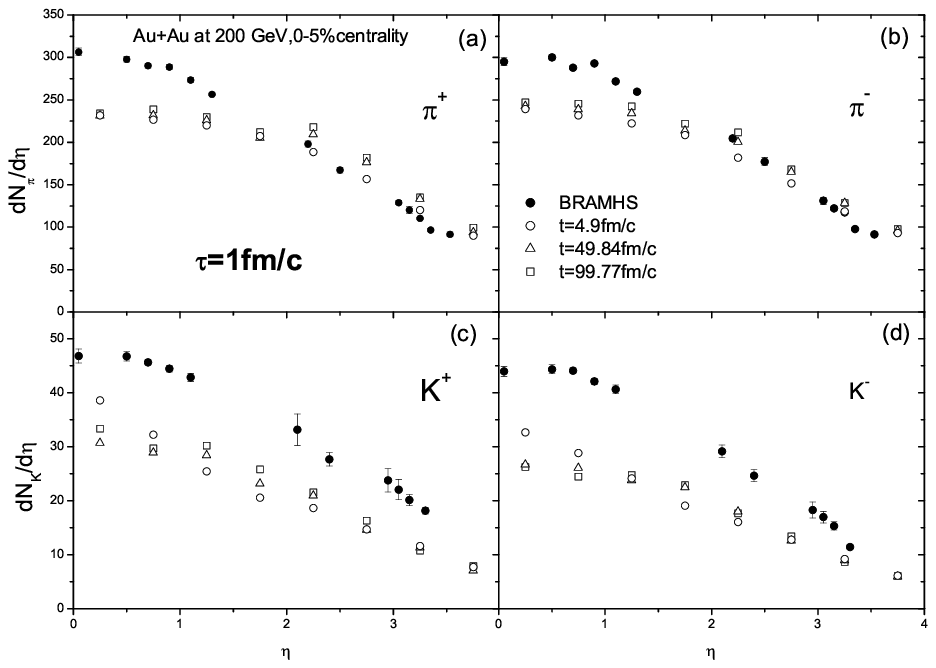}}
\vspace{0.2in} \caption{Particles production  versus
pseudo-rapidity in Au+Au collisions at
 200 GeV C.M. energy at  different  times for proper formation time equal to 1 fm/c.}
\label{fig10}
\end{figure}
\begin{figure}[ht]
\centerline{\hspace{-0.5in}
\includegraphics[width=4.5in,height=3.0in,angle=0]{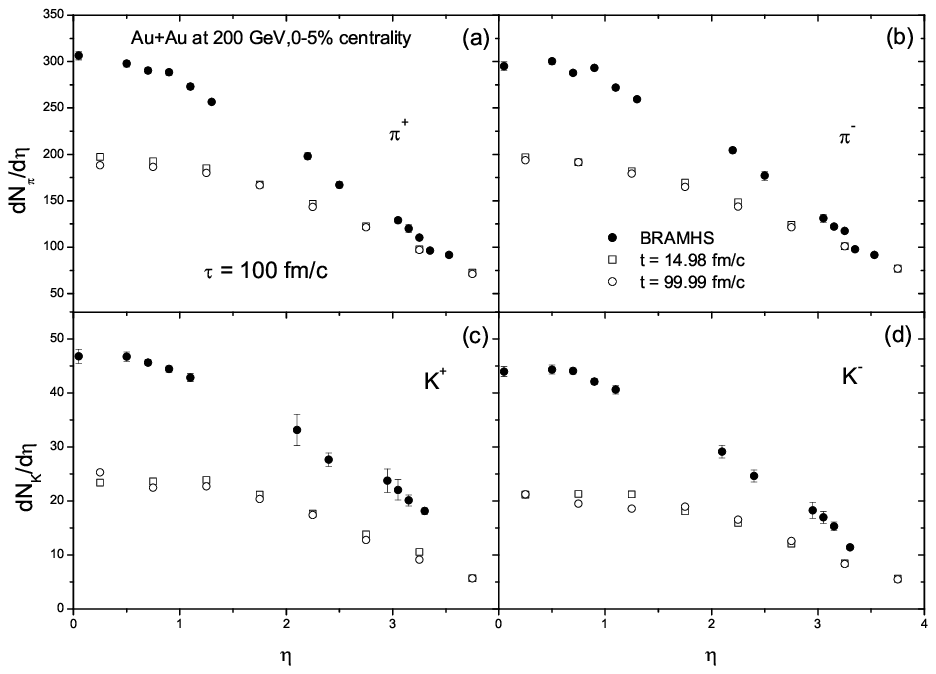}}
\vspace{0.2in} \caption{Particles production  versus
pseudo-rapidity in Au+Au collisions at
 200 GeV C.M. energy at  different  times for proper formation time equal to 100 fm/c.}
\label{fig11}
\end{figure}

\section{Time evolution of the collisions}

Even though the model is still at its infant stage we can discuss
some of its features in more details also to see where and
possibly what we have to do to improve it.  We found that the
model is very sensitive to the input elementary collisions which
we parametrized from very low energy up to the threshold for
Pythia on which we rely for the description of the elementary data
at higher energies.  Our strategy was in any case to fit the
Pythia parameters to pp data when available and then turn to the
AA collisions case.  One
 surprising result for us (but probably known to the authors of \cite{aich}) was that low   $\eta$ particles
come from the final stages of the reaction at  longer times.  This is similar to evaporation from a
compound nucleus where the lowest energy particles (for instance neutrons) are emitted last.  We have
seen already that low  $\eta$ particles come from colliding hadrons having a C.M. energy less than 4 GeV.
 This is also rather surprising to us since we are dealing with a Au+Au collision at 200 GeV and now we
want to see more in detail the various time stages of the reaction.
\begin{figure}[ht]
\centerline{\hspace{-0.5in}
\includegraphics[width=4.5in,height=3.0in,angle=0]{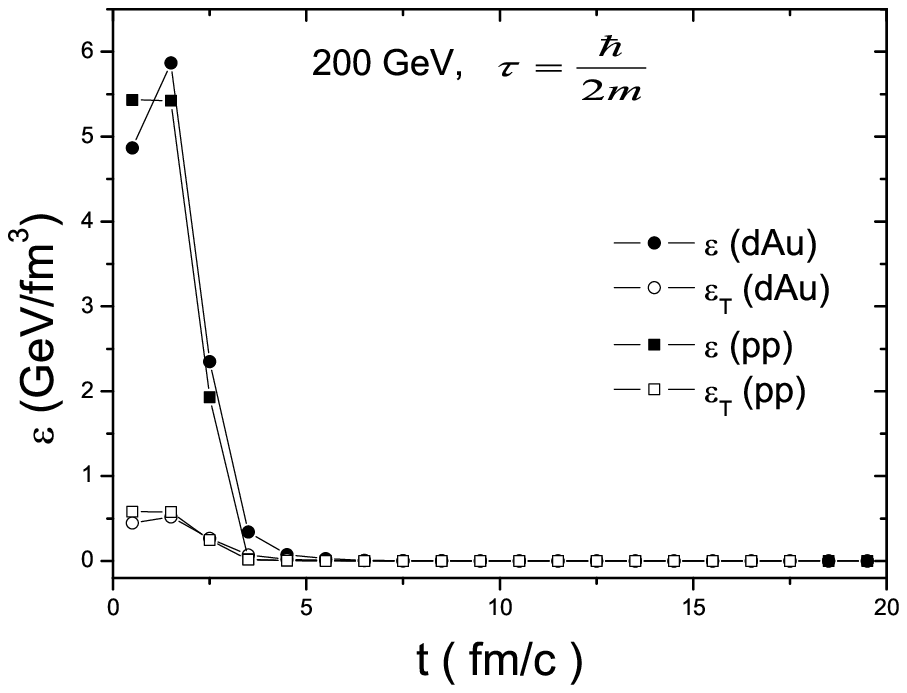}}
\vspace{0.2in} \caption{Energy densities vs. time in a central
cell of radius 2 fm in pp and d+Au collisions at
 200 GeV C.M. and zero impact parameter.}
\label{fig12}
\end{figure}

In figure 9 we display the pseudo-rapidity distribution at various
times of the collision for the 'Heisenberg'
 ht.  We see that the high pseudo-rapidity tail (i.e. $\eta > 5$) saturates already in less than 1 fm/c, while
lower pseudo-rapidipseudo-rapidityty particles are produced at times larger than 5 fm/c. It is obvious to think now that if
these particles are created at such late times, they might not carry any relevant information about the QGP.

In figures 10-11 we display the pseudo-rapidity distribution for a  ht
equal to 1 and 100 fm/c.  The difference with the previous case is striking even though both ht fit the elementary 
pp data.  The effect of the largest ht is to block all secondary collisions, thus the result given in the
figure is coming from first chances collisions only and it is similar to the one of fig. 9 obtained in the first
fm/c of the reaction.  Thus a large ht is equivalent say to Pythia folded for the possible number of 
first chance collisions as calculated in the Eikonal approximation for instance \cite{gei}.

In order to discuss the possibility of QGP formation, its lifetime
and if it reaches equilibrium or not we define a sphere around the
C.M.  of the system of radius 2 fm, similarly to low energy
\cite{aldo} and UrQMD calculations \cite{belk}.  In fact, if a QGP
is formed it should happen in such a sphere first, also if
equilibrium is reached (at least in our model) we should get
isotropy in momentum space at least in  a central region.  In order to
see this we look at the energy density and 'transverse' energy
density in such a central region.  We define $\epsilon_T=\sum
\sqrt{\frac{3}{2}(p^2_x+p^2_y)+m^2}$, where p and m are the
momenta of the particles, x and y the directions perpendicular to
the beam axis z, and the numerical factor is included in order to
have $\epsilon_T=\epsilon$ if equilibrium is reached.  The sum
runs over all the particles within the central sphere.   In figure
(12) we plot the energy densities vs. time for pp and d+Au
collisions.  Two features are worth noticing.  First the
transverse energy density is not much different for the two
systems and second equilibrium is almost reached at times of the
order of 5 fm/c but at low energy density values (less than $0.5
GeV/fm^{3}$).
\begin{figure}[ht]
\centerline{\hspace{-0.5in}
\includegraphics[width=4.5in,height=3.0in,angle=0]{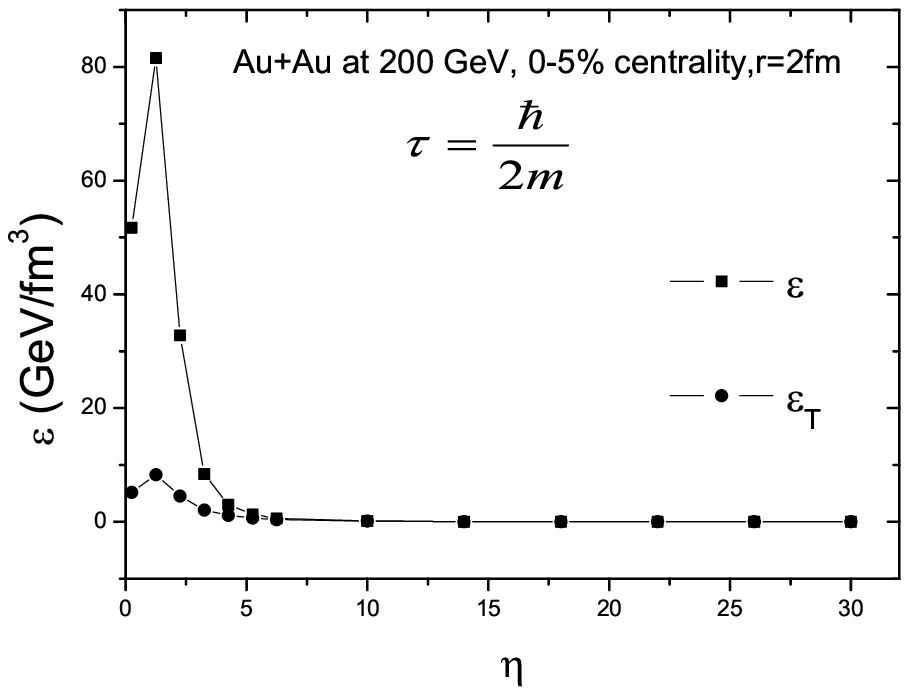}}
\vspace{0.2in} \caption{Same as figure (12) but for Au+Au
collisions.} \label{fig12}
\end{figure}

The situation is dramatically different for Au+Au collisions
(figure 13) at the same RHIC energy.  Now the reached energy
densities are much larger and equilibrium is reached at 4 fm/c
similar to UrQMD calculations \cite{belk}.
 On similar grounds we can calculate how many hadrons and partons are in the central cell at different times,fig.14.
Here the partons are calculated from the number of produced particles at each time and that have lived less
than the ht.

At first the number of hadrons (baryons)  decreases because they
have collided and formed new particles.  After some ht partons
coalesce and are counted as hadrons again.  At this stage they
could collide again and form new particles.  The number of partons
increases quickly with time and it seems to saturate already at 3
fm/c.  There is a stage of non equilibrium (kinetic) but after 4
fm/c the system is equilibrated.  At this time the ratio of
partons to hadrons is about 10 and this could be considered as a
QGP state.  At later times the ratio decreases and we could
consider this as a mixed state of partons and hadrons. Of course
this effect is simply due to the finite ht since no (first order)
phase transition is explicitly included in the model yet.
Nevertheless, the model suggests a mixed phase in the early stages
of the reaction, together with the possibility of reaching thermal
equilibrium. At later times (not plotted),  the number of partons and hadrons 
decreases because of the expansion of the system.  Still some
collisions occur at low energies which produce new hadrons
(especially pions).  This stage is of course not so interesting
but it could be potentially dangerous since it might wash out all
the signals, for instance large fluctuations in D-mesons
productions\cite{terr05}, intermittency etc...,of the plasma.
For completeness, in the same figure we have plotted the total number of partons and hadrons
in the entire space (which never reaches kinetic equilibrium).
  Here one sees that the number of partons remains constant for a relatively long time
 while the number of hadrons increases with time.  This is due to the
interplay of the finite ht, the newly created partons and the hadronization and it is an entirely dynamical process.
\begin{figure}[ht]
\centerline{\hspace{-0.5in}
\includegraphics[width=4.5in,height=3.0in,angle=0]{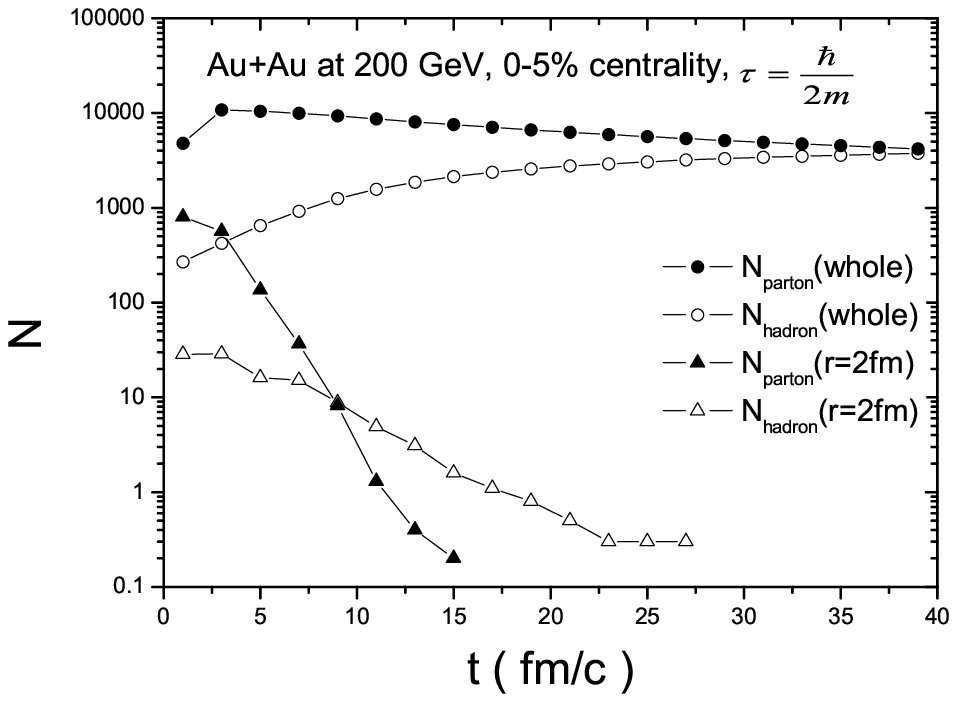}}
\vspace{0.2in} \caption{Hadrons and partons produced versus time
in the central cell for Au+Au collisions.} \label{fig14}
\end{figure}
\section{conclusions}

In this work we have introduced a novel transport approach to
relativistic heavy ion collisions.  Elementary collisions are
chosen following a mean free path method already developed at low
energies.  The elementary process is dealt within the Pythia
approach. The low energies (below 4 GeV C.M. energies) elementary
collisions have been implemented as well.

Within this model we have shown that a purely hadronic approach
fails in reproducing AA data and can reproduce the pp data at a
price of a large (and unreasonable) energy cutoff in the elementary inelastic
collisions.  Finite hadronization times and a careful inclusion of
all possible inelastic channels down to the pion threshold
production gives a good description of pseudo-rapidity
distributions both in pp and Au+Au data.  The possibility of a
finite ht hints to the formation of a mixed QGP and hadronic state
after kinetic equilibrium is reached (i.e. after 5 fm/c).  The
comparison to data in this paper has been restricted to the
highest RHIC energy at the moment.  We plan to extend the
calculations with the present model to other observables and
different energies.  Collective flow calculations are also of
interest. A priority however is the possibility of including a
phase transition in the model to see if the data (or which data)
is sensitive to the phase transition.  This we hope to accomplish
within a Constrained Molecular Model (CoMD) for quarks degree of
freedom recently
 developed \cite{terr04}.

Finally, Z.D.M acknowledges the financial support from  INFN and
Department of Phys. University of Catania in Italy (where most of
the work was performed) and NSFC (10347131) in China. This work is supported in part
 by the EU under
contract CN/ASIA-LINK/008(094-791).  We thank
Prof. Ben-Hao Sa for discussions and for making his code JPCIAE
available to us.

\end{document}